\documentclass[sigconf]{acmart}
\AtBeginDocument{%
  }

\copyrightyear{2026}
\acmYear{2026}
\setcopyright{cc}
\setcctype{by}
\acmConference[UIST Adjunct '26]{The 39th Annual ACM Symposium on User Interface Software and Technology}{November 02--05, 2026}{Detroit, MI, USA}
\acmBooktitle{The 39th Annual ACM Symposium on User Interface Software and Technology (UIST Adjunct '26), November 02--05, 2026, Detroit, MI, USA}
\acmDOI{10.1145/3830397.3841887}
\acmISBN{979-8-4007-2855-6/2026/11}

\begin{document}

\title{The Interface of Theseus: The Rise of Just-In-Time Interfaces}

\author{Michael S. Bernstein}
\email{msb@cs.stanford.edu}
\orcid{0000-0001-8020-9434}
\affiliation{%
  \institution{Stanford University}
  \city{Stanford}
  \state{CA}
  \country{USA}
}

\renewcommand{\shortauthors}{Bernstein}

% \begin{abstract}
% When Your Agent Knows You Better Than You Know Yourself

% Interactive software places power in the hands of the user, on the assumption that the user is the one who knows what they need. But, with access to rich reasoning capabilities and diverse information about our lives, AI agents may soon have a more accurate sense of your needs and goals than you do yourself. This situation enables positive visions such as Maes's agents that reduce work or Apple's Knowledge Navigator. It also leads to ethically fraught visions such as nanny agents. But perhaps less appreciated, it also may deeply reshape software. What happens when an application can ask your agent what you're trying to do? Or if a particular tool would be helpful to surface? Or how to best help you hold to a long-term goal?
% \end{abstract}

\begin{abstract}
Our community creates interfaces---pre-constructed, static artifacts. These interfaces are effortful to create, so we invest great effort and care into each of them. Yet the equilibrium is shifting: we can now construct nearly any  software, bespoke, within moments. When construction is cheap, we may soon live in a future in which interfaces are constructed in real time. This is a world of just-in-time interfaces: rapid, discardable, on-demand interfaces tailored to our needs in the moment. Will our community still create traditional interactive systems in this world? Or will we create interface generators, ``interface.md'' guides for AI consumption? I argue that we ought to lean into a future of rapid, bespoke, on-demand interfaces.
\end{abstract}




%\received{June 10 2026}

\maketitle

Interfaces today are strikingly static artifacts: each interface is designed, engineered, and pre-fabricated to support whatever experiences the creator decided to support. The content rendered through that interface is dynamic, of course, but the interface itself remains stable. In her keynote at CHI 2012, YouTube Director of User Experience Margaret Gould Stewart described this situation as designing a dinner plate: it can't change, even as the food placed on it does change~\cite{stewart2012keynote}. We have sometimes struggled at the bonds of static interfaces by creating adaptive ones~\cite{gajos2010supple}; yet the resulting lack of predictability~\cite{gajos2008predictability} has typically led us back to tried-and-true static interfaces.

But as the speed and cost of software construction falls to astonishingly low levels, what used to be an expensive design and development process will soon happen in mere moments. While our community focuses on AI agents, the bigger tidal wave may in fact be \emph{just-in-time interfaces}: entire interactive systems constructed at the moment of need and customized to exactly the user's situation. Just-in-time interfaces combine two ingredients: (1)~rapid coding agents, and (2)~user models with a strong notion of what we need~\cite{lam2026objectives,shaikh2025gum,maes1994agents,apple1987navigator,cao2025generative}. By combining these ingredients, nearly any situation can be analyzed for what might help and a resulting interaction can be constructed bespoke within moments. Already, AI chatbots render on-demand visualizations and artifacts as we work with them. Soon, why not our entire experience?

Suppose you are trying to purchase your next laptop. As you search, the tradeoffs get materialized into a faceted browsing interface. You ask about differences in performance; the system creates a side-by-side simulation of the two laptops and how smoothly your favorite game will render. It applies its knowledge of you and your workflow to show how your behavior and satisfaction might shift under each of these laptops---maybe you start leaving the larger one at home because it's so heavy or doesn't fit in your purse. Each tool, experience, and visualization is built bespoke for the task. If something that you need is not there, merely ask and it shall appear. Then, when you are done, tear the interface away like a used sheet of paper, and start your next task on a blank one.

This vision of just-in-time interaction leverages realtime induction of the user's objectives~\cite{lam2026objectives}, combined with a strong background model of who the user is and what they need and want~\cite{shaikh2025gum} and agentic coding tools. Many such tools already exist (e.g.,~\cite{cao2025generative, gajos2010supple, chen2026generative, leviathan2026generativeui, min2025meridian, min2026gradual}), and will only improve.

What is the role of our community in such a world~\cite{vaithilingam2024imagining}? When the systems that we create are now mere commodities that we toss out after a few moments?

One reaction is that we will become the meta-architects: those who craft the generators. Following the pattern of agents.md files, Google Labs has released a design.md spec \cite{google2026designmd}: a structured description of a design system that agents use to ensure that the resulting designs all cohere. We might likewise craft interface.md files or other agentic harnesses that guide a system in what to create, when, and how.

Another reaction is one of skepticism: that this will never work. The idea of just-in-time interaction certainly has its critics \cite{okopnyi2024against}. Prior attempts at adaptive interaction largely failed with users; Microsoft famously rolled back its adaptive menus after users complained that the system was unpredictable \cite{harris2006personalizedmenus}. Just-in-time interfaces inherently violate many principles of interaction design, especially  Nielsen's heuristics of consistency and standards \cite{nielsen1994heuristics}---what we might term, more simply, familiarity. The result kills any opportunity to build systems that we gain implicit familiarity with and that eventually become ready-to-hand \cite{winograd1986understanding}. While the ever-changing staircases of Hogwarts make for amazing spectacle, they do not make for a particularly useful way of getting where you're going.

Yet there may be undeniable benefits that overcome these shortcomings. Joel Spolsky once wrote that, according to traditional usability principles, peer-to-peer file sharing systems such as Napster never should have worked: they have terrible affordances and were incredibly confusing to use \cite{spolsky2004usability}. But, he argued, they gained widespread adoption because they were \emph{useful}---so useful that we overlooked their shortcomings. A future of bespoke, just-in-time interaction offers some opportunities unmatched by our traditionally architected designs: for example, they can integrate information across the boundaries of my traditional walled garden applications. Already, systems can gather this data through MCP or by simple screenshot monitoring \cite{shaikh2025gum}. Suddenly, a bespoke interface can connect the project I'm discussing on Slack with the code I'm writing and the todo list I've been maintaining in order to proactively create a mission control system for completing my goals.

Ultimately, I argue that we will soon experience an equilibrium shift: when software construction becomes effectively free and instantaneous, then interfaces will become temporary. We labored under static interfaces when they were difficult to create. Now, our role will be more akin to a toolkit creator: we create the levers that shape what gets created and how they operate, but we may never lay hands or eyes on the last-mile interfaces themselves.

\bibliographystyle{ACM-Reference-Format}
\bibliography{interface-of-theseus}


\end{document}